  [2003/10/13 v1 Conversano 2003 write-up driver file (Philip G. Ratcliffe)]
\documentclass{wqcd03}                 

\usepackage{txfonts}                   

\usepackage{Ratcliffe}

\confname{%
  QCD@Work 2003 - International Workshop on QCD,
  Conversano, Italy, 14--18 June 2003
}

\title{\acs{QCD} and Transverse-Spin Physics}

\author{Philip G. Ratcliffe}

\address{%
  Dipartimento di Scienze CC.FF.MM.\\
  Universit{\`a} degli Studi dell'Insubria---sede di Como\\
  \emph{and}\\
  Istituto Nazionale di Fisica Nucleare---sezione di Milano
}

\begin{document}

\begin{abstract}
  A pedagogical presentation of \aclp{SSA} and transversity is offered.
  Detailed discussion is given of various aspects of single-spin asymmetries in
  lepton--nucleon and in hadron--hadron scattering and of the r{\^o}le of
  \acl{pQCD} and evolution in the context of transversity.
\end{abstract}

\maketitle
\begin{fmffile}{fmffigs}
\section{Preamble}

Let me begin by remarking that many of the topics touched on here are covered
in much greater depth in \cite{Barone:2001sp,Barone:2003fy}. Therefore, much
credit and thanks are due to my two collaborators Barone and Drago. Further,
less condensed and more complete reviews than the present may also be found in
other recent proceedings: see, \eg, \cite{Ratcliffe:2002qb.0}.

By way of motivation for the subject area, I should open by noting that the
general theoretical framework for discussing transverse-spin effects, at least
at a basic level, is now rather solid. Added to which, on the experimental side
there is presently a great deal of activity: witness the programmes of HERMES
at DESY, COMPASS at CERN and the spin programme at RHIC. It is, however, also
true that much theoretical work is still necessary to unravel the
phenomenology; both to perform serious and relevant future data analysis and to
indicate which measurements could be most usefully performed.

What is \emph{transverse} spin? By ``transverse'' one means that the spin vector
is perpendicular to the particle momentum (\cf parallel or longitudinal, as in
the talk by \citet{Ridolfi:0000tp}). This terminology should not be confused
with the traditional designation ``transverse state'', as applied to the case of
gauge bosons (where the \acs{EM} fields lie in the transverse plane): left- and
right-handed circular polarisations correspond to helicity states and therefore
to a \emph{longitudinal} spin vector. It is also important to stress that
transverse polarisation itself does \emph{not} depend on particle masses (\cf
the natural polarisation of the LEP beam). However, the problem of mass can and
does arise when seeking \emph{measurable} transverse-spin effects, which almost
inevitably require spin flip.

\subsection{Transversity}

Transversity, which simply describes the probability of finding a quark
polarised parallel (as opposed to antiparallel) to a
\emph{transversely}-polarised parent hadron, has a rather long history: the
concept (though \emph{not} the term) was introduced in \cite{Ralston:1979ys}
via Drell--Yan processes; its \ac[\hyphenate]{LO} anomalous dimensions were
first calculated in \cite{Baldracchini:1981uq} but, unfortunately, languished
\emph{forgotten} for over a decade! They were recalculated much later in
\cite{Artru:1990zv} and, still early on, \emph{unwittingly} (as part of the
evolution of $g_2$), by a number of authors
\cite{Kodaira:1979ib,Antoniadis:1981dg,Bukhvostov:1985rn,Ratcliffe:1986mp}.

\subsection{The \acs{DIS} structure function \boldmath{$g_2$}}

Dubbed ``\emph{the nucleon's other spin dependent structure function}'' by
\citet{Jaffe:1994cj}, the \ac{DIS} structure function $g_2$ has an even longer
history. Already in \citeyear{Hey:1972pp} its scaling behaviour had been
examined \cite{Hey:1972pp,Heimann:1973hq,Ahmed:1975tj}; as already noted, the
\ac{LO} evolution in \acs{QCD} was calculated by various authors
\cite{Kodaira:1979ib,Antoniadis:1981dg,Bukhvostov:1985rn,Ratcliffe:1986mp}
(although incorrectly in the earlier papers).

Now, It is important to appreciate here that $g_2$ is \emph{very} different to
the better-known $F_2$ and $g_1$ \ac{DIS} structure functions: it is
essentially twist-three and therefore involves three-parton correlators; it can
thus have \emph{no} partonic interpretation. While it is true that in the
Wandzura--Wilczek approximation $g_2$ may be related to $g_1$
\cite{Wandzura:1977qf}, this is only via explicit neglect of the higher-twist
contributions and it is now largely accepted that there is no compelling reason
for so doing.

\subsection{\Aclp{SSA}}

\Acp{SSA} perhaps represent the \emph{oldest} form of high-energy spin
measurement: the only requirement is either a polarised beam or target (and for
$\Lambda^0$ production \emph{neither} is necessary). However, after early
interest (due to the surprisingly large magnitudes found experimentally), a
theoretical \emph{dark age} descended on \acp{SSA}: apparently \ac{pQCD} had
nothing to say, save that they ought to \emph{vanish}. We now realise that the
rich phenomenology is matched by a richness of the theoretical framework. This
will, in essence, be a central theme of the present talk.

Before continuing I have to admit that one might argue that the $Q^2$ of
existing \ac{SSA} data is too low for \ac{pQCD} to be applicable. Indeed, there
are many \emph{non}-\acs{pQCD} models that explain part (but never all) of the
data; Some examples may be found in
\cite{Andersson:1979wj,DeGrand:1981pe,Barni:1992qn,Soffer:1992am}. Here,
however, I shall examine \acp{SSA} purely within the \ac{pQCD} framework. It is
also true that the present data show \emph{no} indication that such effects are
dying off with growing $Q^2$.

\section{Introduction}

\subsection{\Aclp{SSA}}

Generically, \acp{SSA} reflect correlations of the form
\begin{equation}
  \vec{s} \cdot \left( \vec{p} \vprod \vec{k} \right) ,
  \label{eq:triple-product}
\end{equation}
where $\vec{s}$ is a spin vector, $\vec{p}$ and $\vec{k}$ are particle/jet
momenta. Indeed, it should not be difficult to convince oneself that the
constraint of parity conservation imposes such a form when only one spin vector
is available. A typical example might be: $\vec{s}$ a target polarisation
vector (transverse), $\vec{p}$ the beam direction and $\vec{k}$ a final-state
particle direction. Therefore, polarisations involved in \acp{SSA} must
typically be transverse with respect to the reaction plane, although there are
exceptions.

Transforming basis from transverse spin to helicity via
\begin{equation}
  \Ket{\uparrow/\downarrow} =
  \tfrac1{\sqrtnob2\,} \left[ \strut\, \Ket{+} \pm \I\Ket{-} \right] ,
\end{equation}
any such asymmetry takes on the (schematic) form
\begin{equation}
  \mathcal{A}_N
  \sim
  \frac{\braket{\uparrow|\uparrow}-\braket{\downarrow|\downarrow}}
       {\braket{\uparrow|\uparrow}+\braket{\downarrow|\downarrow}}
  \sim
  \frac{2\Im\braket{+|-}}{\braket{+|+}+\braket{-|-}} \, .
\end{equation}
The form of the second numerator indicates \emph{interference} between
amplitudes, where one is \emph{spin-flip} and the other \emph{non-flip}, with a
relative \emph{phase difference}.

It was soon realised \cite{Kane:1978nd} that a gauge theory such as \acs{QCD}
in the Born approximation and massless (or high-energy) limit cannot satisfy
either requirement: fermion helicity is conserved and tree diagrams are real.
This provoked the statement \cite{Kane:1978nd} that ``\dots\ \emph{observation
of significant polarizations in \emph{[pion production]} would contradict
either \acs{QCD} or its applicability.}'' Clearly, however, \acs{QCD} is still
alive and well, despite a large number of sizable, measured single
transverse-spin effects.

It was not long, however, before \citet*{Efremov:1982sh.0} opened up an escape
route via consideration of the three-parton correlators involved in, \eg,
$g_2$: they demonstrated that the relevant mass scale for helicity flip is not
the current quark mass, but a typical hadronic mass and that the
pseudo-two-loop nature of the diagrams can lead to an imaginary part in certain
regions of partonic phase space. Unfortunately, quite some time passed before
the richness of the available structures was recognised and brought fully to
fruition, see \cite{Qiu:1991pp.0}.

\subsection{Transversity}

Transversity is the third (and \emph{final}) twist-two partonic distribution
function. At this point it is important to make the distinction between
partonic distributions (or densities) (\eg, $q(x)$, $\DL{q}(x)$, $\DT{q}(x)$,
\dots) and \ac{DIS} structure functions ($F_1$, $F_2$, $g_1$, $g_2$, \dots). In
the unpolarised and helicity-dependent cases at leading twist there is a
simple, rather direct, correspondence between the two: \ac{DIS} structure
functions are just weighted sums of parton densities. However, as already
noted, in the case of transverse-spin: (\emph{i}) there is no \ac{DIS}
transversity structure function and (\emph{ii}) $g_2$ cannot be expressed in
terms of a partonic densities.

\begin{figure}[hbtp]
  \centering
  \includegraphics
    [width=0.20\textwidth,bb=158 571 310 681,clip]{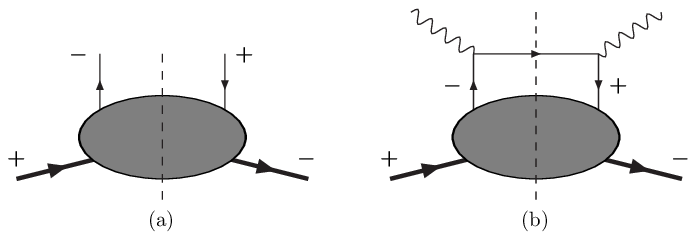}
  \hfil
  \includegraphics
    [width=0.20\textwidth,bb=336 571 486 679,clip]{epsfiles/chirality}
  \caption{%
    (a) The chirally-odd hadron--quark amplitude for $h_1$ and
    (b) the \emph{forbidden} chirality-flip \acs{DIS} handbag diagram.
  }
  \label{fig:chirality}
\end{figure}

The three twist-two structures are then
\begin{subequations}
\begin{align}
  q(x) &=
  \int \! \frac{\D\xi^-}{4\pi} \, \E^{\I xP^+ \xi^-} \!
  \bra{PS}
    \anti\psi(0)
      \gamma^+
    \psi(0, \xi^-\!, \Vec{0}_\perp)
  \ket{PS} ,
\\
  \DL{q}(x) &=
  \int \! \frac{\D\xi^-}{4\pi} \, \E^{\I xP^+ \xi^-} \!
  \bra{PS}
    \anti\psi(0)
      \gamma^+ \gamma_5
    \psi(0, \xi^-\!, \Vec{0}_\perp)
  \ket{PS} ,
\\
  \DT{q}(x) &=
  \int \! \frac{\D\xi^-}{4\pi} \, \E^{\I xP^+ \xi^-} \!
  \bra{PS}
    \anti\psi(0)
      \gamma^+ \gamma^1 \gamma_5
    \psi(0, \xi^-\!, \Vec{0}_\perp)
  \ket{PS} .
\end{align}
\end{subequations}
The presence of the $\gamma_5$ matrix signals generic spin dependence while the
$\gamma^1$ in $\DT{q}(x)$ signals helicity flip, \emph{precluding} transversity
contributions in \ac{DIS}, see Fig.~\ref{fig:chirality}. N.B. chirality flip is
not a problem if the quarks connect to different hadrons, as in the \ac{DY}
process.

\subsection{\Acl{pQCD} evolution}

\begin{figure}[hbtp]
  \centering
  \begin{fmfgraph*}(20,25)
    \fmfpen{thin}
    \fmfset{curly_len}{2mm}
    \fmfbottom{i1,o2}
    \fmftop{o1,i2}
    \fmf{fermion,label=$+$,l.side=left}{i1,v1}
    \fmf{fermion,label=$+$,l.side=left}{v1,o1}
    \fmf{fermion,label=$-$,l.side=left}{i2,v2}
    \fmf{fermion,label=$-$,l.side=left}{v2,o2}
    \fmffreeze
    \fmf{gluon}{v1,v2}
  \end{fmfgraph*}
  \hfil\hfil
  \begin{fmfgraph*}(20,25)
    \fmfpen{thin}
    \fmfset{curly_len}{2mm}
    \fmfbottom{i1,o2}
    \fmftop{o1,i2}
    \fmf{fermion}{i1,v1}
    \fmf{gluon}{o1,v1}
    \fmf{gluon}{v2,i2}
    \fmf{fermion}{v2,o2}
    \fmffreeze
    \fmf{fermion,label=$+$,l.side=left}{i1,v1}
    \fmf{fermion,label=$-$,l.side=left}{v2,o2}
    \fmf{fermion,label=?}{v1,v2}
  \end{fmfgraph*}
  \caption{%
    Left, the evolution kernel in a physical (axial) gauge for transversity;
    right, an excluded gluon--fermion mixing diagram.
  }%
  \label{fig:rungs}
\end{figure}
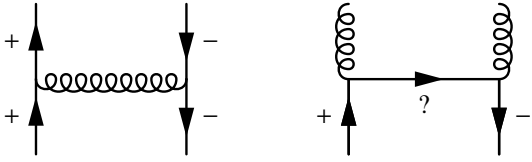

The non-diagonal nature of transversity in a helicity basis forces diagonality
in flavour space, see Fig.~\ref{fig:rungs},
and thus the \ac{LO} \acs{QCD} evolution of transversity is of the non-singlet
type. The quark line cannot return to the same hadronic blob and therefore
there can be neither quark--gluon mixing nor mixing between different types of
quark.

\begin{figure}[hbtp]
  \centering
  \psfrag{0.5}{\mbox{}}
  \psfrag{1.5}{\mbox{}}
  \psfrag{2.5}{\mbox{}}
  \psfrag{3.5}{\mbox{}}
  \psfrag{(a)}{(a)}
  \psfrag{(b)}{(b)}
  \psfrag{x}{$x$}
  \psfrag{u}{\tiny$\;u$}
  \psfrag{D}{\tiny$\Delta$}
  \psfrag{T}{\tiny$\;_T$}
  \psfrag{LO input}{\tiny LO input}
  \psfrag{NLO input}{\tiny \quad NLO input}
  \psfrag{LO evolution}{\tiny LO evolution}
  \psfrag{NLO evolution}{\tiny NLO evolution}
  \includegraphics[width=0.45\textwidth]{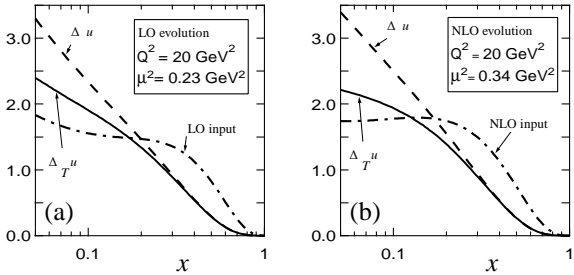}
  \caption{%
    A comparison of the $Q^2$-evolution of $\DT{u}(x,Q^2)$ and $\DL{u}(x,Q^2)$ at
    (a) \acs{LO} and (b) \acs{NLO}; from \cite{Hayashigaki:1997dn}.
  }%
  \label{fig:hkk-fig6}
\end{figure}

The \ac{LO} non-singlet DGLAP quark--quark splitting functions are:
\begin{subequations}
\begin{align}
  P_{qq}^{(0)}
  &=
  \CF \left( \frac{1+x^2}{1-x\,} \right)_+ ,
\\
  \DL{P}_{qq}^{(0)}
  &=
  P_{qq}^{(0)}
  \qquad \mbox{(helicity conservation)} ,
\\
  \DT{P}_{qq}^{(0)}
  &=
  \CF \left[ \left( \frac{1+x^2}{1-x\,} \right)_+ - 1 + x \right] .
\end{align}
\end{subequations}
One sees that while the first moments of $P_{qq}^{(0)}$ and $\DL{P}_{qq}^{(0)}$
both \emph{vanish} (leading to well-known conservation laws and sum rules), the
same does \emph{not} hold for $\DT{P}_{qq}^{(0)}$. The overall effect is a
decrease in transversity with respect to helicity densities as $Q^2$ increases,
see Fig.~\ref{fig:hkk-fig6}.

\subsection{The \citeauthor{Soffer:1995ww} bound}
\def\phast{{\vphantom{*}}}

\begin{figure}[hbtp]
  \centering
  \includegraphics[width=0.2\textwidth]{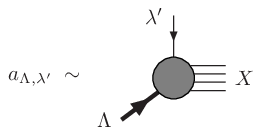}
  \caption{Hadron--parton helicity amplitudes, see \cite{Soffer:1995ww}.}
  \label{fig:hadron-quark}
\end{figure}

\citet{Soffer:1995ww} has derived an interesting and non-trivial bound
involving all three leading-twist structures. In terms of hadron--quark
helicity amplitudes, see Fig.~\ref{fig:hadron-quark},
the quark densities may be expressed as
\begin{subequations}
\begin{align}
  q(x) & \propto
  \Im (\mathcal{A}_{++,++} +\mathcal{A}_{+-,+-})
  \propto
  \sum_X ( a_{++}^* a_{++}^\phast + a_{+-}^* a_{+-}^\phast ) \, ,
  \label{soffer0.3}
\\
  \DL{q}(x) & \propto
  \Im (\mathcal{A}_{++,++} -\mathcal{A}_{+-,+-})
  \propto
  \sum_X ( a_{++}^* a_{++}^\phast - a_{+-}^* a_{+-}^\phast ) \, ,
  \label{soffer0.4}
\\
  \DT{q}(x) & \propto
  \Im \mathcal{A}_{+-,-+}
  \propto
  \sum_X a_{--}^* a_{++}^\phast \, .
  \label{soffer0.5}
\end{align}
\end{subequations}
The following Schwartz identity (combined with parity conservation)
\begin{equation}
  \sum_X |a_{++}^\phast \pm a_{--}^\phast|^2 \ge 0
  \; \Rightarrow \;
  \sum_X a_{++}^* a_{++}^\phast \pm \sum_X a_{--}^* a_{++}^\phast \ge 0 \, ,
\end{equation}
then leads to
\begin{equation}
  q_+(x)\ge|\DT{q}(x)|
  \quad\mbox{or}\quad
  q(x)+\DL{q}(x)\ge2|\DT{q}(x)| \, .
\end{equation}

Note that, while saturation of the bound is, of course, not necessarily
expected \apriori, it is rather suggestive that the physical magnitude of
$\DT{q}(x)$ might well be intermediate to $q(x)$ and $\DL{q}(x)$. Indeed, there
are many arguments for expecting that $\DT{q}(x)$ should be of a similar
strength to $\DL{q}(x)$ (for example, see \cite{Barone:2001sp}), at least at
some sufficiently low energy scale.

\subsection{A \acs{DIS} definition for transversity}

Quark density functions find their natural definition in the lepton--nucleon
\ac{DIS} process, where the parton model is usually formulated and
non-perturbative models developed. On translation to \ac{DY}, it is well known
that large $K$ factors of $\text{O}(\pi\alpha_s)$ appear. At RHIC energies this
represents an approximately $30\%$ correction while at EMC/SMC energies it is
nearly $100\%$. As is well known, such corrections are indeed corroborated by
the data.

\begin{figure}[hbtp]
  \centering
  \begin{fmfgraph*}(40,30)
    \fmfpen{thick}
    \fmfleft{i1,i2}
    \fmfright{o1,o2}
    \fmf{fermion,tension=1.0}{i1,u1,v1}
    \fmf{fermion,tension=0.5}{v1,v2}
    \fmf{fermion,tension=1.0}{v2,u2,o1}
    \fmf{photon}{i2,v1}
    \fmf{scalar}{v2,o2}
    \fmffreeze
    \fmf{gluon}{u1,u2}
    \fmfv{label=\rnode{NA}{},l.a=0}{v2}
  \end{fmfgraph*}
  \caption{A \acs{DIS} Higgs--photon interference diagram.}
  \label{fig:higgs1}
\end{figure}
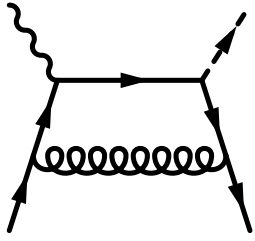

The pure \ac{DY} coefficient functions are known for transversity, see
\cite{Vogelsang:1993jn,Contogouris:1994ws,Kamal:1996as,Vogelsang:1998ak}, but
are scheme dependent. Moreover, a term $\ln^2x/(1-x)$ appears, which is
\emph{not} found for spin-averaged \cite{Altarelli:1979ub} or
helicity-dependent \cite{Ratcliffe:1983yj} \ac{DY}. Added to problems arising
with a vector--scalar current product \cite{Blumlein:2001ca}, This suggests
that an interesting check is in order. In order to have a \acs{DIS}-like
process as a starting point, it is clearly necessary to allow for helicity flip
somewhere. This may be most conveniently achieved via the introduction of a
scalar vertex, see Fig.~\ref{fig:higgs1}.
A certain amount of care is needed as an extra contribution from the scalar
vertex must be absorbed into the running mass (or Higgs-like coupling
constant).

Armed with such a process, in the standard manner one may now calculate a
coefficient function for transversity in \ac{DIS}, which combined with the
already-known coefficient for \ac{DY} will provide the corresponding $K$
factor. The three cases of unpolarised, longitudinally  and transversely
polarised are displayed in the following equations:
\begin{subequations}
\begin{align}
  C^f_{q,\text{DY}} - 2 C^f_{q,\text{DIS}}
  &=
  \frac{\alpha_s}{2\pi} \, \CF
  \left[
    \left(\frac43\pi^2+1\right)\delta(1-x)
    +
    \frac3{(1-x)_+}
  \right .
  \nonumber
\\
  &\hspace*{2.5em} \left. \null
    + 2(1+x^2)\left(\frac{\ln(1-x)}{1-x}\right)_+
    - 6 - 4x
  \right] ,
\\
  C^g_{q,\text{DY}}-2C^g_{q,\rm DIS}
  &=
  C^f_{q,\text{DY}}-2C^f_{q,\rm DIS} +
  \frac{\alpha_s}{2\pi} \, \CF
  \left[ \strut
    2 + 2x
  \right] ,
\\
  C^h_{q,\text{DY}} - 2 C^h_{q,\text{DIS}}
  &=
  \frac{\alpha_s}{2\pi} \, \CF
  \left[
    \left(\frac43\pi^2-1\right)\delta(1-x)
    +
    \frac{3x}{(1-x)_+}
  \right.
  \nonumber
\\
  &\hspace*{0.75em} \left. \null
    - 6x\frac{\ln^2x}{1-x}
    + 4 - 4x
    + 4x\left(\frac{\ln(1-x)}{1-x}\right)_+
  \right] ,
\end{align}
\end{subequations}
where $\CF=\frac43$ is the just usual colour-group Casimir for the fermion
representation. The small difference in the coefficient of the
$\delta$-function is not actually significant, the most striking difference is
the appearance, already mentioned, of the $\ln^2x/(1-x)$ term. We note that
while one might object that any substantial differences are probably due to the
peculiar \ac{DIS} definition adopted for transversity, this term arises in the
\ac{DY} calculation and has its origin in the particular phase-space
integration required by the fixing of the final lepton-pair azimuthal angle.

\subsection{\acs{DIS}--\acs{DY} transversity asymmetry}

\begin{figure}[hbtp]
  \centering
  \psfrag{A_DY}{$A_\mathrm{DY}$}
  \psfrag{tau}{$\tau$}
  \includegraphics[width=0.45\textwidth]{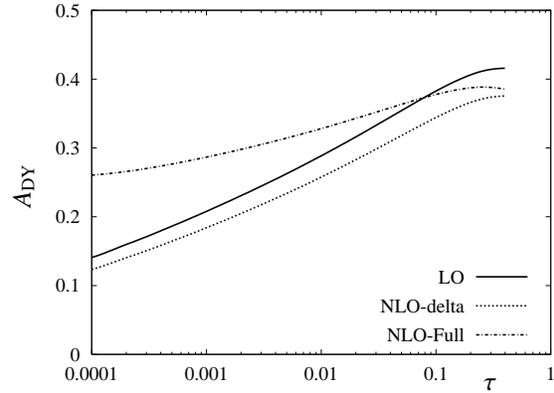}
  \caption{%
    The transversity asymmetry (\emph{valence quarks only}) for Drell--Yan. The
    variables are $\tau=Q^2/s$, $s=4{\cdot}10^4\,\GeV^2$, with kinematic limits
    $\tau<x1,x2<1$; see the text for a description of the different curves.
  }
  \label{fig:dyasym}
\end{figure}

Using the above results it is now possible to evaluate the effect of such a $K$
factor on the \ac{DY} transversity asymmetry. In Fig.~\ref{fig:dyasym} we
display the asymmetry $A_\mathrm{DY}$ for transversely polarised protons in the
\ac{DY} process.
The full curve shows the \ac[\hyphenate]{LO} case while the dotted line shows
the effect of the $\delta$-function contribution and the dot--dashed line
represents the full calculation. One can clearly see that, in contrast to the
helicity case \cite{Ratcliffe:1983yj}, the terms beyond the $\delta$-function
alter the behaviour quite considerably. This is a signal that direct comparison
with model calculations might not be as straightforward as hoped.

One might, of course, argue that this is just an artifact of the peculiar
\ac{DIS} definition used. However, the substantial departures from the familiar
behaviour occur in the \ac{DY} calculation (and are traceable to the phase
space alterations due to the requirement of \emph{not} integrating over the
azimuthal angle of the lepton pair). In any case, work is under way to perform
similar calculations for the various possible combinations of scalar and vector
currents in \ac{DIS} and \ac{DY}, in order to confirm the origins of the large
$K$ factor found.

\section{Single-Hadron Production}

\begin{figure}[hbtp]
  \centering
  \includegraphics[width=0.4\textwidth]{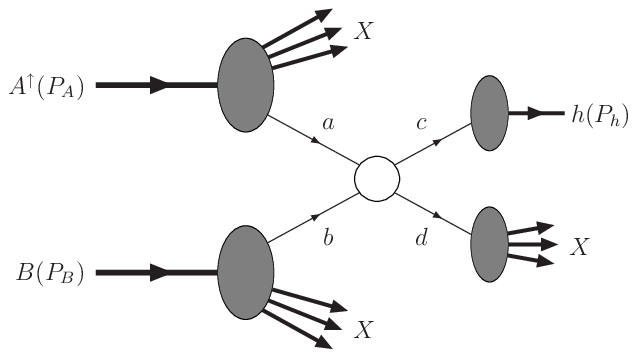}
  \caption{%
    Hadron--hadron scattering with a single polarised initial-state hadron
    ($A^\uparrow$).
  }
  \label{fig:single}
\end{figure}

While the cleanest and most unambiguous experimental access to transversity
should nevertheless lie in the \ac{DY} process, \acp{SSA} represent a more
immediately available (if not necessarily accessible) source of information.
Thus, I shall now briefly examine single-hadron production off a transversely
polarised target:
\begin{equation}
  A^\uparrow(P_A) \, + \, B(P_B) \, \to \, h(P_h) \, + \, X \, .
\end{equation}
Hadron $A$ is transversely polarised and the unpolarised (or spinless) hadron
$h$ (which may also be a photon) is produced at \emph{large} transverse
momentum $\Vec{P}_{hT}$, thus \ac{pQCD} is applicable. The process is shown
pictorially in Fig.~\ref{fig:single}.

In a typical experimental example $A$ and $B$ are protons while $h$ is a pion.
One measures an \acs{SSA}:
\begin{equation}
  A_{T}^h =
  \frac{\D\sigma(\Vec{S}_T) - \D\sigma(-\Vec{S}_T)}
       {\D\sigma(\Vec{S}_T) + \D\sigma(-\Vec{S}_T)} \, .
\end{equation}

\begin{figure}[hbtp]
  \centering
  \includegraphics[width=0.3\textwidth]{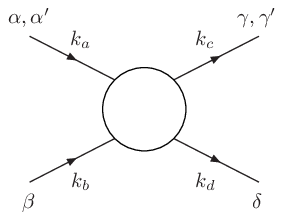}
  \caption{%
    The hard-scattering parton amplitude
    $\mathcal{M}_{\alpha\beta\gamma\delta}$.
  }
  \label{fig:partonamp}
\end{figure}

According to the factorisation theorem, the differential cross-section for the
reaction may be written formally as
\begin{equation}
  \D\sigma =
  \sum_{abc} \sum_{\alpha\alpha'\gamma\gamma'}
  \rho^a_{\alpha'\alpha} \,
  f_a(x_a) \otimes
  f_b(x_b) \otimes
  \D\hat\sigma_{\alpha\alpha'\gamma\gamma'} \otimes
  \mathcal{D}_{h/c}^{\gamma'\gamma}(z) \, .
\end{equation}
Here $f_a$ ($f_b$) is the density of parton $a$ ($b$) in hadron $A$ ($B$),
$\rho^a_{\alpha\alpha'}$ is the spin density matrix of parton $a$,
$\mathcal{D}_{h/c}^{\gamma\gamma'}$ is the fragmentation matrix of parton $c$
into hadron $h$ and $\D\hat\sigma/\D\hat{t}$ is the elementary cross-section:
\begin{equation}
  \left(
    \frac{\D\hat\sigma}{\D\hat{t}}
  \right)_{\alpha\alpha'\gamma\gamma'}
  =
  \frac1{16\pi\hat{s}^2} \, \frac12 \, \sum_{\beta\delta}
  \mathcal{M}_{\alpha\beta\gamma\delta} \,
  \mathcal{M}^*_{\alpha'\beta\gamma'\delta} \, ,
\end{equation}
where $\mathcal{M}_{\alpha\beta\gamma\delta}$ is the amplitude for the hard
partonic process, shown in Fig.~\ref{fig:partonamp}.

For an unpolarised produced hadron, the off-diagonal elements of
$\mathcal{D}_{h/c}^{\gamma\gamma'}$ vanish, \ie,
$\mathcal{D}_{h/c}^{\gamma\gamma'}\propto\delta_{\gamma\gamma'}$. Then,
helicity conservation implies $\alpha=\alpha'$ and there can be no dependence
on the spin of hadron $A$ and thus all \acp{SSA} are identically zero. Such a
conclusion, in stark contrast with reality, may be avoided by considering
either intrinsic quark transverse motion, or higher-twist effects.

\subsection{Transverse motion and \acsp{SSA}}

Quark intrinsic transverse motion can generate \acp{SSA} in three different
ways:
\bgroup\parskip0pt%
\begin{enumerate} \itemsep=0pt \parskip=0pt
\item[1.]
$\Vec\kappa_T$ in hadron $h$ implies $\mathcal{D}_{h/c}^{\gamma\gamma'}$ may be
non-diagonal ($T$-odd effect at the fragmentation level).
\item[2.]
$\Vec{k}_T$ in hadron $A$ requires $f_a(x_a)$ to be replaced by
$\mathcal{P}_a(x_a,\Vec{k}_T)$, which may depend on the spin of $A$ ($T$-odd
effect at the distribution level).
\item[3.]
$\Vec{k}'_T$ in hadron $B$ requires $f_b(x_b)$ to be replaced by
$\mathcal{P}_b(x_b,\Vec{k}'_T)$. The transverse spin of $b$ in the unpolarised
$B$ may then couple to the transverse spin of $a$ ($T$-odd effect at the
distribution level).
\end{enumerate}
\egroup%
The three mechanisms are, correspondingly:
\bgroup\parskip0pt%
\begin{enumerate} \itemsep=0pt \parskip=0pt
\item[1.]
the \citeauthor{Collins:1993kk} effect \cite{Collins:1993kk};
\item[2.]
the \citeauthor{Sivers:1990cc} effect \cite{Sivers:1990cc};
\item[3.]
an effect in Drell--Yan studied by \citet{Boer:1999mm}.
\end{enumerate}
\egroup%
Note that all such intrinsic-$\Vec\kappa_T$, -$\Vec{k}_T$, or -$\Vec{k}'_T$
effects are $T$-odd and therefore they require initial- or final-state
interactions. When quark transverse motion is included, the \acs{QCD}
factorisation theorem is \emph{not} proven.

Assuming factorisation to be valid, the cross-section is
\begin{align}
  E_h \, \frac{\D^3\sigma}{\D^3\Vec{P}_h} &=
  \sum_{abc} \sum_{\alpha\alpha'\beta\beta'\gamma\gamma'}
  \frac1{\pi z}
  \nonumber
\\
  & \quad \null \times
  \int \! \D{x}_a
  \int \! \D{x}_b
  \int \! \D^2\Vec{k}_T
  \int \! \D^2\Vec{k}'_T
  \int \! \D^2\Vec\kappa_T
  \nonumber
\\
  & \quad \null \times
  \mathcal{P}_a(x_a, \Vec{k}_T) \, \rho^a_{\alpha'\alpha} \,
  \mathcal{P}_b(x_b, \Vec{k}'_T) \, \rho^b_{\beta'\beta}
  \nonumber
\\
  & \quad \null \times
  \left(
    \frac{\D\hat\sigma}{\D\hat{t}}
  \right)_{\alpha\alpha'\beta\beta'\gamma\gamma'}
  \mathcal{D}_{h/c}^{\gamma'\gamma}(z, \Vec\kappa_T) \, ,
\end{align}
where
\begin{equation}
  \left(
    \frac{\D\hat\sigma}{\D\hat{t}}
  \right)_{\alpha\alpha'\beta\beta'\gamma\gamma'}
  =
  \frac1{16\pi\hat{s}^2} \, \sum_{\beta\delta}
  \mathcal{M}_{\alpha\beta\gamma\delta}
  \mathcal{M}^*_{\alpha'\beta\gamma'\delta} \, .
\end{equation}

The Collins mechanism requires that we take into account the intrinsic quark
transverse motion inside the produced hadron $h$, and neglect the transverse
momenta of all other quarks (assuming the spin of $A$ to be directed along
$y$):
\begin{align}
  E_h \,&\frac{\D^3\sigma( \Vec{S}_T)}{\D^3\Vec{P}_h} -
  E_h \, \frac{\D^3\sigma(-\Vec{S}_T)}{\D^3\Vec{P}_h}
  \nonumber
\\
  & \null =
  -2 \, | \Vec{S}_T | \, \sum_{abc}
  \int \! \D{x}_a
  \int \! \D{x}_b
  \int \! \D^2\Vec\kappa_T \, \frac1{\pi z}
  \nonumber
\\
  & \quad \null \times
  \DT{f}_a(x_a) \, f_b(x_b) \,
  \Delta_{TT} \hat\sigma(x_a, x_b, \Vec\kappa_T) \,
  \DT^0 D_{h/c}(z, \Vec\kappa_T^2) \, ,
\end{align}
where, $\Delta_{TT}\hat\sigma$ is a partonic spin-transfer asymmetry.

The Sivers effect relies on $T$-odd distribution functions and predicts a
single-spin asymmetry of the form
\begin{align}
  E_h \,&\frac{\D^3\sigma( \Vec{S}_T)}{\D^3\Vec{P}_h} -
  E_h \, \frac{\D^3\sigma(-\Vec{S}_T)}{\D^3\Vec{P}_h}
  \nonumber
\\
  & \null =
  | \Vec{S}_T | \, \sum_{abc}
  \int \! \D{x}_a
  \int \! \D{x}_b
  \int \! \D^2\Vec{k}_T \, \frac1{\pi z}
  \nonumber
\\
  & \quad \null \times
  \Delta_0^T f_a(x_a, \Vec{k}_T^2) \, f_b(x_b) \,
  \frac{\D\hat\sigma(x_a, x_b, \Vec{k}_T)}{\D\hat{t}} \, D_{h/c}(z) \, ,
\end{align}
where $\Delta_0^T{f}$ (related to $f_{1T}^\perp$) is a $T$-odd distribution.

Finally, the effect studied in \cite{Boer:1999mm} gives rise to an asymmetry
involving the other $T$-odd distribution, $\DT^0f$ (related to $h_1^\perp$):
\begin{align}
  E_h \,&\frac{\D^3\sigma( \Vec{S}_T)}{\D^3\Vec{P}_h} -
  E_h \, \frac{\D^3\sigma(-\Vec{S}_T)}{\D^3\Vec{P}_h}
  \nonumber
\\
  & \null = -
  2 | \Vec{S}_T | \, \sum_{abc}
  \int \! \D{x}_a
  \int \! \D{x}_b
  \int \! \D^2\Vec{k}'_T \, \frac1{\pi z}
  \nonumber
\\
  & \quad \null \times
  \DT{f}_a(x_a)\, \DT^0 f_b(x_b, \Vec{k'}_T^2) \,
  \Delta_{TT} \hat\sigma'(x_a, x_b, \Vec{k}'_T) \, D_{h/c}(z) \, ,
\end{align}
where $\Delta_{TT}\hat\sigma'$ is the partonic initial-state spin-correlation
asymmetry.

\subsection{Higher-twist and \acsp{SSA}}

As already mentioned, it was first pointed out in \cite{Efremov:1982sh.0} that
non-vanishing \acp{SSA} can also be generated in \ac{pQCD} by resorting to
higher twist and the so-called gluonic poles present in diagrams involving
$qqg$ correlators. Such asymmetries were later evaluated in the context of
\acs{QCD} factorisation in \cite{Qiu:1991pp.0}, where direct photon production
was studied and, more recently, hadron production \cite{Qiu:1998ia}. This
program has now been further extended to cover the chirally-odd contributions
in \cite{Kanazawa:2000hz.0}.

The possibilities multiply when higher-twist is taken into consideration as the
new contribution can reside in any one of the three building blocks: parton
densities, hard-scattering processes or fragmentation functions. Thus one has
the following general expression:
\begin{align}
  \D\sigma &=
  \sum_{abc}
  \left\{ \vphantom{D_{h/c}^{(3)}}
    G_F^a(x_a, y_a) \otimes f_b(x_b) \otimes
    \D\hat\sigma \otimes D_{h/c}(z)
  \right.
  \nonumber
\\
  & \hspace{2.5em} \null +
  \DT{f}_a(x_a) \otimes E_F^b(x_b, y_b) \otimes
  \D\hat\sigma' \otimes D_{h/c}(z)
  \nonumber
\\[1ex]
  & \hspace{2.5em} \null +
  \left.
    \DT{f}_a(x_a) \otimes f_b(x_b) \otimes
    \D\hat\sigma'' \otimes D_{h/c}^{(3)}(z)
  \right\} .
\end{align}
The first term does not contain transversity and is the chirally-even mechanism
studied in \cite{Qiu:1998ia}; the second is the chirally-odd contribution
analysed in \cite{Kanazawa:2000hz.0}; and the third contains a twist-three
fragmentation function $D_{h/c}^{(3)}$.

\subsection{Phenomenology}

\citet{Anselmino:2002pd.0} have compared the data to various models for
partonic densities based on the previous possible ($k_T$) contributions and
find good descriptions. However, they cannot yet differentiate between
contributions. The higher-twist calculations of \citet*{Qiu:1991pp.0} are
rather opaque, involving many diagrams, complicated momentum flow, colour and
spin structure. The twist-three correlators (as found in $g_2$) obey
constraining relations with $k_T$-dependent densities; thus, in fact, the two
approaches are not entirely independent. Indeed, it is well known that
\cite{Politzer:1974fr} that higher-twist may always be traded off for non-zero
$k_T$.

\section{A Novel Factorisation}

\begin{figure}[hbtp]
  \centering
  \includegraphics[width=0.25\textwidth,bb=92 558 196 701,clip]
                  {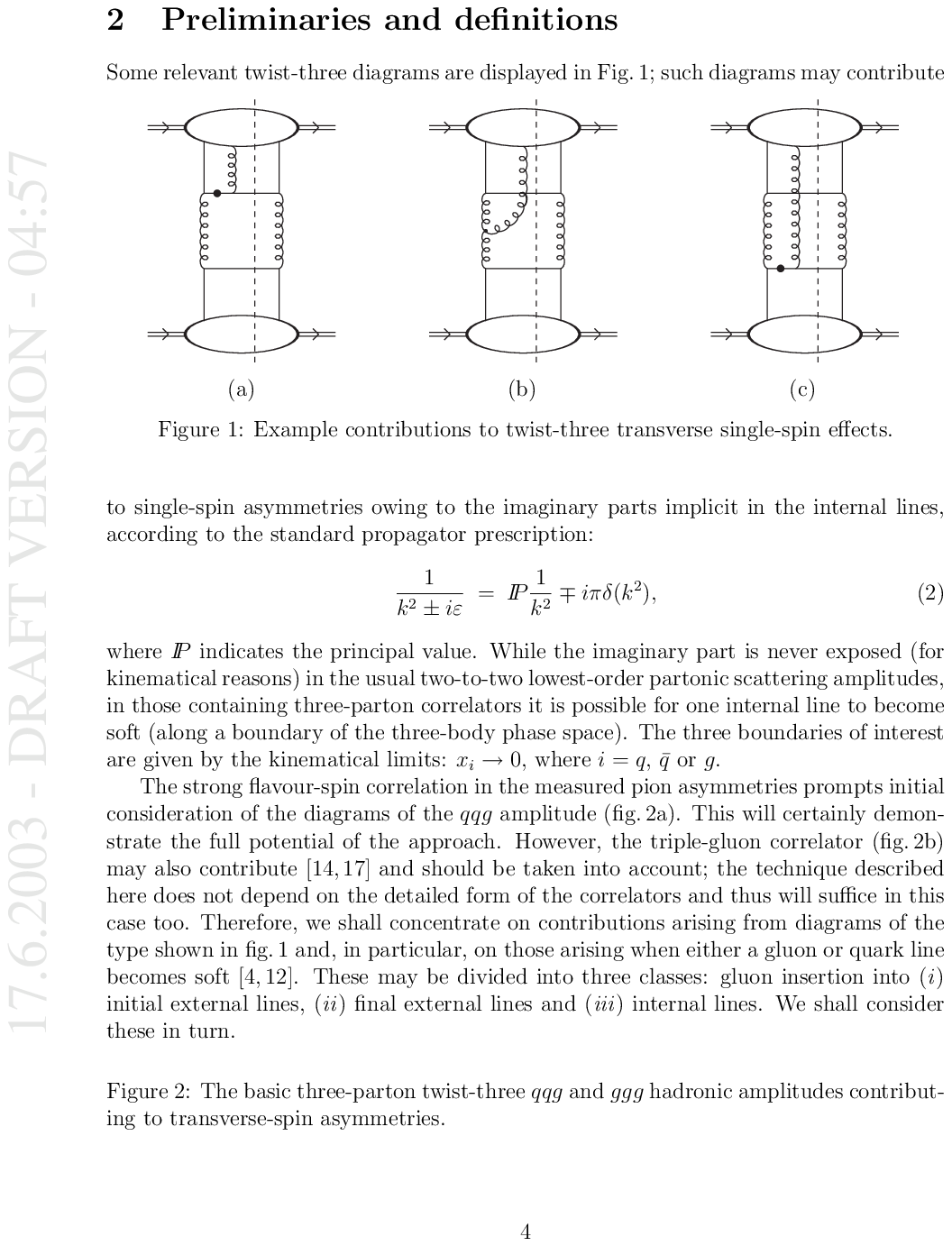}
  \caption{%
    A typical twist-three diagram giving rise to pole terms.
  }
  \label{fig:pole-terms}
\end{figure}

The manner in which twist-three diagrams involving three-parton correlators
(such as in Fig.~\ref{fig:pole-terms})
can supply an imaginary part via a pole term (spin-flip is implicit in the
particular operators considered, those relating to $g_2$ in \ac{DIS}) is as
follows \cite{Efremov:1982sh.0}: the standard propagator prescription,
\begin{equation}
  \frac1{k^2\pm{\I}\varepsilon} =
  \mathrm{I\!P} \frac1{k^2} \mp \I\pi\delta(k^2) \, ,
\end{equation}
then leads to an imaginary contribution for $k^2\to0$. Thus, mechanisms
generating unsuppressed \acp{SSA} may easily be constructed. This is precisely
the origin of the contributions successfully exploited in \cite{Qiu:1991pp.0}
to obtain large \acp{SSA}.

\subsection{Pole Diagrams}
\begin{figure}[hbtp]
  \centering
  \includegraphics[height=23mm,bb=142 661 250 749,clip]
                  {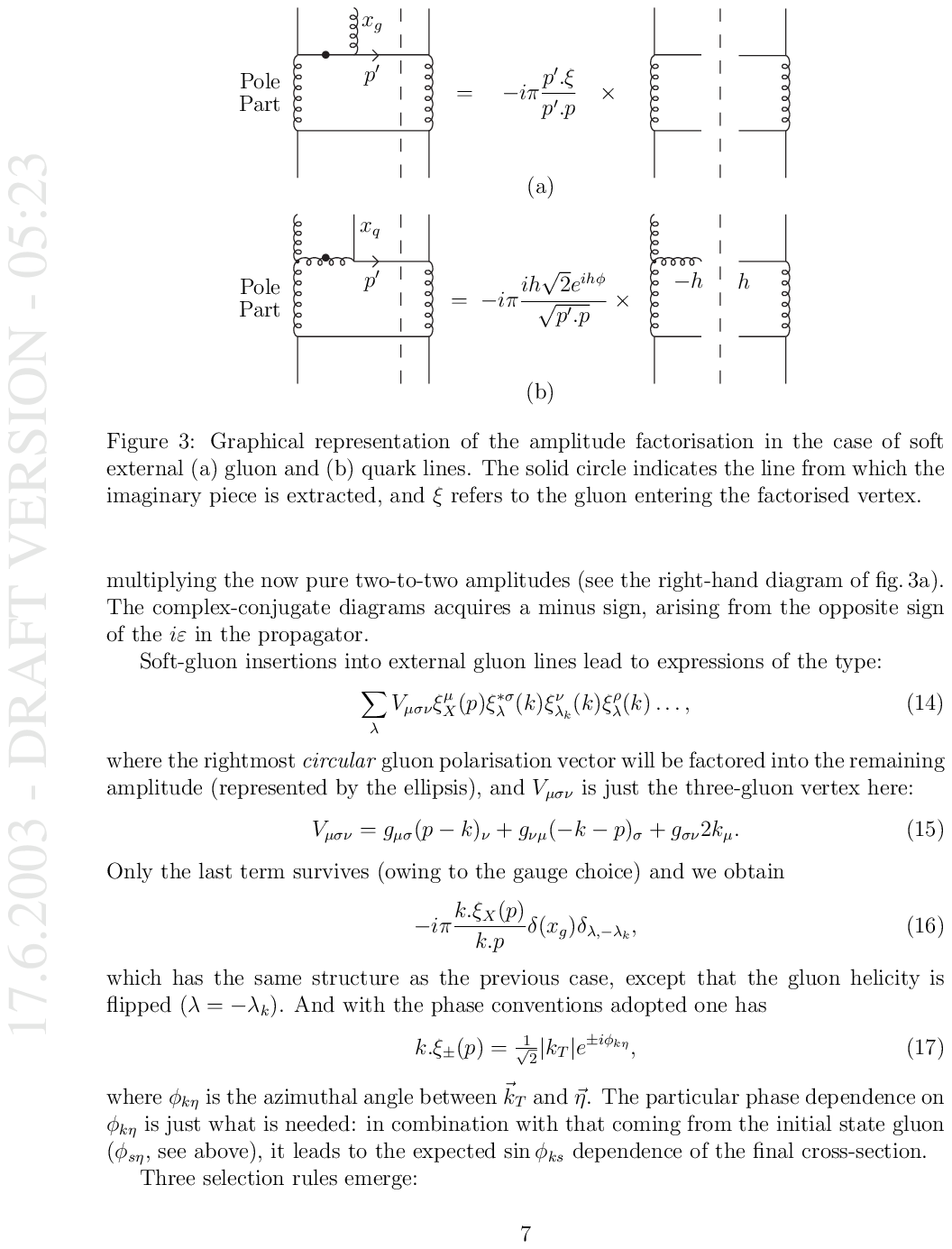}
  \includegraphics[height=23mm,bb=250 661 437 749,clip]
                  {epsfiles/pole-factor}
  \caption{%
    A graphical representation of the factorisation of the pole term.
  }
  \label{fig:pole-factor}
\end{figure}

The peculiar kinematics of the configuration involved may be further exploited
to factorise the amplitude in a rather convenient and suggestive manner
\cite{Ratcliffe:1998pq}. For a gluon $x_gp$ inserted into an (initial or final)
external line $p'$, $k=p'-x_gp$ and this means $x_g\to0$. The result is
represented graphically in Fig.~\ref{fig:pole-factor}.
The term $p'{\cdot}\xi$ contains the typical triple product involving the spin vector
shown in Eq.~\ref{eq:triple-product}. The obvious interpretation is then that
three-parton amplitudes in general may be factorised into corresponding
two-parton amplitudes multiplied by simple kinematical factors, in which all
the spin information resides. Such factorisation can be performed
systematically for all poles (gluon and fermion), \emph{i.e.}, on all external
legs with all possible soft insertions. This still generates rather complex
structures: there are many possible such insertions for any given correlator,
with contributions of different signs and momentum dependence.

\subsection{Large-\boldmath{$\Nc$}}

\begin{figure}[hbtp]
  \centering
  \includegraphics[width=0.25\textwidth,bb=391 558 495 701,clip]
                  {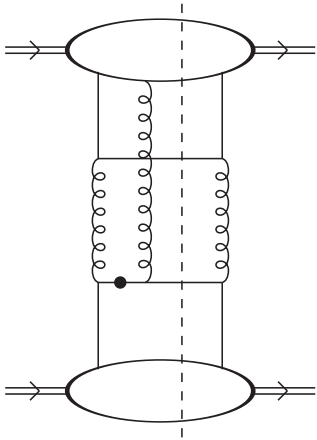}
  \caption{%
    The single pole-term diagram surviving in the large-$\Nc$ limit.
  }
  \label{fig:pole-terms-bis}
\end{figure}

The structure of these twist-three contributions may be still further
simplified by considering the colour structure of the various diagrams
involved, which is also very different. In all cases (examined) it turns out
that just one diagram, shown in Fig.~\ref{fig:pole-terms-bis}, dominates in the
large-$\Nc$ limit.
All other possible insertions (leading to an imaginary part via a pole) are
suppressed by $1/\Nc^2$. Note that this is true irrespective of whether to
final-state partons are quarks or gluons. At this point it becomes rather
trivial to estimate (with the approximations made) the size and sign of
\acp{SSA} that may be generated via such mechanisms in any given hadronic
process.

We should note in concluding this section that the whole procedure still needs
to be repeated for all the other twist-three contributions (\emph{e.g.}, also
in fragmentation). However, the simplifications demonstrated obviously render
all such mechanisms very transparent, allowing easy and rapid evaluation of
their impact in \ac{SSA}.

Finally, for lack of time, I have not commented on and can only mention other
recent developments concerning, for example, non-standard time reversal
\cite{Anselmino:2001vn}, final-state interactions \cite{Brodsky:2002cx} and
non-trivial Wilson lines \cite{Collins:2002kn}.

\section{Conclusions and Outlook}

Single-spin asymmetries have passed from a dark age, during which there was
essentially \emph{no} (\acs{QCD}-based) theory, to a period of illumination,
where there is almost \emph{too much}. Hopefully, the multiplicity of
contributions can be reduced to a few simple terms:
\begin{itemize} \itemsep=1ex \parskip=0ex
\item
experiment can eliminate some possibilities if null results are obtained;
\item
relationships between three-parton correlators and $k_T$-dependent densities
should show the equivalence between phenomenological models;
\item
pole-factorisation and the large-$\Nc$ limit should simplify calculations and
allow a simple pattern to emerge.
\end{itemize}
Fortunately, the experimental activity, now on the increase, is matched by a
continued interest in the phenomenology on the part of a long-standing group of
spin theorists. This is thus certainly an area of hadronic physics that is
destined to produce interesting (and perhaps) surprising results in the not too
distant future.

Somewhat in contrast, the case of transversity is rather cut-and-dried as far
as the theoretical interpretation is concerned. What is now lacking is
experimental knowledge of this spin density. Such data will complete our
understanding of the spin structure of the proton. Moreover, the peculiar
nature of the \ac{pQCD} evolution of transversity (it is of the pure
non-singlet form) could, in principle, allow interesting studies of scaling
violations.

\end{fmffile}

\begin{thebibliography}{}
  \makeatletter
    \newtoks\@addcite
    \def\addcite#1{\nocite{#1}\csname b@#1\@extra@b@citeb\endcsname}
    \def\additem#1#2{\@addcite={#2}\immediate\write\@auxout
      {\string\@newl@bel{b}{#1\@extra@b@citeb}{\the\@addcite}}}
    \immediate\write\@auxout{\string\@ifundefined{etal}%
      {\string\gdef\string\etal{\string\emph{et~al.}}}{}}%
    \immediate\write\@auxout{\string\@ifundefined{citeauthoryear}%
      {\string\gdef\string\citeauthoryear\string#1\string#2\string#3%
        {\string#1, \string#2, \string#3}}{}}%
  \makeatother
  \providecommand{\citeauthoryear}[3]{#1, #2, #3}
  \providecommand{\etal}{\emph{et~al.}}
  \providecommand{\href}[2]{#2}
  \providecommand{\ibid}{\emph{ibid.}\ }
  \providecommand{\addcite}{Addendum/erratum~\cite}
  \providecommand{\noopsort}[1]{}
  \providecommand{\volumeface}[1]{\textup{\textbf{#1}}}

\bibitem[Barone \etal(2002)Barone, Drago and Ratcliffe]{Barone:2001sp}
V.~Barone, A.~Drago and P.G.~Ratcliffe, \emph{Phys. Rep.} \textbf{359}
  (2002)~1;
\href{http://www.arXiv.org/abs/hep-ph/0104283}
  {\mbox{\textsf{hep-ph/0104283}}}.

\bibitem[Barone \etal(2003)Barone and Ratcliffe]{Barone:2003fy}
V.~Barone and P.G.~Ratcliffe,
\href{http://www.worldscientific.com.sg/books/physics/5052.html}
  {\emph{Transverse Spin Physics}} (World Sci., 2003).

\bibitem[Ratcliffe(2003)Ratcliffe]{Ratcliffe:2002qb.0}
P.G.~Ratcliffe, in Proc. of the \emph{Advanced Study Institute on Symmetries
  and Spin---{Praha}--{SPIN} 2002} (Prague, July 2002);
\emph{Czech. J. Phys. Suppl.} (2003) to appear;
\href{http://www.arXiv.org/abs/hep-ph/0211222}
  {\mbox{\textsf{hep-ph/0211222}}};
\addcite{Ratcliffe:2002xv.0}.

\bibitem[Ridolfi(9999)Ridolfi]{Ridolfi:0000tp}
G.~Ridolfi, these proceedings.

\bibitem[Ralston \etal(1979)Ralston and Soper]{Ralston:1979ys}
J.~Ralston and D.E.~Soper, \emph{Nucl. Phys.} \textbf{B152} (1979)~109.

\bibitem[Baldracchini \etal(1981)Baldracchini \etal]{Baldracchini:1981uq}
F.~Baldracchini, N.S.~Craigie, V.~Roberto and M.~Socolovsky, \emph{Fortschr.
  Phys.} \textbf{30} (1981)~505.

\bibitem[Artru \etal(1990)Artru and Mekhfi]{Artru:1990zv}
X.~Artru and M.~Mekhfi, \emph{Z. Phys.} \textbf{C45} (1990)~669.

\bibitem[Kodaira \etal(1979)Kodaira \etal]{Kodaira:1979ib}
J.~Kodaira, S.~Matsuda, K.~Sasaki and T.~Uematsu, \emph{Nucl. Phys.}
  \textbf{B159} (1979)~99.

\bibitem[Antoniadis \etal(1981)Antoniadis and Kounnas]{Antoniadis:1981dg}
I.~Antoniadis and C.~Kounnas, \emph{Phys. Rev.} \textbf{D24} (1981)~505.

\bibitem[Bukhvostov \etal(1985)Bukhvostov \etal]{Bukhvostov:1985rn}
A.P.~Bukhvostov, {\'E}.A.~Kuraev, L.N.~Lipatov and G.V.~Frolov, \emph{Nucl.
  Phys.} \textbf{B258} (1985)~601.

\bibitem[Ratcliffe(1986)Ratcliffe]{Ratcliffe:1986mp}
P.G.~Ratcliffe, \emph{Nucl. Phys.} \textbf{B264} (1986)~493.

\bibitem[Jaffe(1995)Jaffe]{Jaffe:1994cj}
R.L.~Jaffe, in Proc. of the \emph{{CERN} {SMC} Meeting on the Internal Spin
  Structure of the Nucleon} (New Haven, Jan. 1994), eds. V.W.~Hughes and
  C.~Cavata (World Sci., 1995), p.~110.

\bibitem[Hey \etal(1972)Hey and Mandula]{Hey:1972pp}
A.J.G.~Hey and J.E.~Mandula, \emph{Phys. Rev.} \textbf{D5} (1972)~2610.

\bibitem[Heimann(1973)Heimann]{Heimann:1973hq}
R.L.~Heimann, \emph{Nucl. Phys.} \textbf{B64} (1973)~429.

\bibitem[Ahmed \etal(1975)Ahmed and Ross]{Ahmed:1975tj}
M.A.~Ahmed and G.G.~Ross, \emph{Phys. Lett.} \textbf{B56} (1975)~385.

\bibitem[Wandzura \etal(1977)Wandzura and Wilczek]{Wandzura:1977qf}
S.~Wandzura and F.~Wilczek, \emph{Phys. Lett.} \textbf{B72} (1977)~195.

\bibitem[Andersson \etal(1979)Andersson, Gustafson and
  Ingelman]{Andersson:1979wj}
B.~Andersson, G.~Gustafson and G.~Ingelman, \emph{Phys. Lett.} \textbf{B85}
  (1979)~417.

\bibitem[DeGrand \etal(1981)DeGrand and Miettinen]{DeGrand:1981pe}
T.A.~DeGrand and H.I.~Miettinen, \emph{Phys. Rev.} \textbf{D24} (1981)~2419;
\addcite{DeGrand:1981pe.e}.

\bibitem[Barni \etal(1992)Barni, Preparata and Ratcliffe]{Barni:1992qn}
R.~Barni, G.~Preparata and P.G.~Ratcliffe, \emph{Phys. Lett.} \textbf{B296}
  (1992)~251.

\bibitem[Soffer \etal(1992)Soffer and Tornqvist]{Soffer:1992am}
J.~Soffer and N.A.~Tornqvist, \emph{Phys. Rev. Lett.} \textbf{68} (1992)~907.

\bibitem[Kane \etal(1978)Kane, Pumplin and Repko]{Kane:1978nd}
G.L.~Kane, J.~Pumplin and W.~Repko, \emph{Phys. Rev. Lett.} \textbf{41}
  (1978)~1689.

\bibitem[Efremov \etal(1982)Efremov and Teryaev]{Efremov:1982sh.0}
A.V.~Efremov and O.V.~Teryaev, \emph{Yad. Fiz.} \textbf{36} (1982)~242;
\addcite{Efremov:1982sh.t}; \addcite{Efremov:1985ip.0}.

\bibitem[Qiu \etal(1991)Qiu and Sterman]{Qiu:1991pp.0}
J.-W.~Qiu and G.~Sterman, \emph{Phys. Rev. Lett.} \textbf{67} (1991)~2264;
\addcite{Qiu:1992wg.0}.

\bibitem[Hayashigaki \etal(1997)Hayashigaki, Kanazawa and
  Koike]{Hayashigaki:1997dn}
A.~Hayashigaki, Y.~Kanazawa and Y.~Koike, \emph{Phys. Rev.} \textbf{D56}
  (1997)~7350;
\href{http://www.arXiv.org/abs/hep-ph/9707208}
  {\mbox{\textsf{hep-ph/9707208}}}.

\bibitem[Soffer(1995)Soffer]{Soffer:1995ww}
J.~Soffer, \emph{Phys. Rev. Lett.} \textbf{74} (1995)~1292;
\href{http://www.arXiv.org/abs/hep-ph/9409254}
  {\mbox{\textsf{hep-ph/9409254}}}.

\bibitem[Vogelsang \etal(1993)Vogelsang and Weber]{Vogelsang:1993jn}
W.~Vogelsang and A.~Weber, \emph{Phys. Rev.} \textbf{D48} (1993)~2073.

\bibitem[Contogouris \etal(1994)Contogouris, Kamal and
  Merebashvili]{Contogouris:1994ws}
A.P.~Contogouris, B.~Kamal and Z.~Merebashvili, \emph{Phys. Lett.}
  \textbf{B337} (1994)~169.

\bibitem[Kamal(1996)Kamal]{Kamal:1996as}
B.~Kamal, \emph{Phys. Rev.} \textbf{D53} (1996)~1142;
\href{http://www.arXiv.org/abs/hep-ph/9511217}
  {\mbox{\textsf{hep-ph/9511217}}}.

\bibitem[Vogelsang(1998)Vogelsang]{Vogelsang:1998ak}
W.~Vogelsang, \emph{Phys. Rev.} \textbf{D57} (1998)~1886;
\href{http://www.arXiv.org/abs/hep-ph/9706511}
  {\mbox{\textsf{hep-ph/9706511}}}.

\bibitem[Altarelli \etal(1979)Altarelli, Ellis and
  Martinelli]{Altarelli:1979ub}
G.~Altarelli, R.K.~Ellis and G.~Martinelli, \emph{Nucl. Phys.} \textbf{B157}
  (1979)~461.

\bibitem[Ratcliffe(1983)Ratcliffe]{Ratcliffe:1983yj}
P.G.~Ratcliffe, \emph{Nucl. Phys.} \textbf{B223} (1983)~45.

\bibitem[Bl{\"u}mlein(2001)Bl{\"u}mlein]{Blumlein:2001ca}
J.~Bl{\"u}mlein, \emph{Eur. Phys. J.} \textbf{C20} (2001)~683;
\href{http://www.arXiv.org/abs/hep-ph/0104099}
  {\mbox{\textsf{hep-ph/0104099}}}.

\bibitem[Collins(1993)Collins]{Collins:1993kk}
J.C.~Collins, \emph{Nucl. Phys.} \textbf{B396} (1993)~161;
\href{http://www.arXiv.org/abs/hep-ph/9208213}
  {\mbox{\textsf{hep-ph/9208213}}}.

\bibitem[Sivers(1990)Sivers]{Sivers:1990cc}
D.~Sivers, \emph{Phys. Rev.} \textbf{D41} (1990)~83.

\bibitem[Boer(1999)Boer]{Boer:1999mm}
D.~Boer, \emph{Phys. Rev.} \textbf{D60} (1999) 014012;
\href{http://www.arXiv.org/abs/hep-ph/9902255}
  {\mbox{\textsf{hep-ph/9902255}}}.

\bibitem[Qiu \etal(1999)Qiu and Sterman]{Qiu:1998ia}
J.-W.~Qiu and G.~Sterman, \emph{Phys. Rev.} \textbf{D59} (1999) 014004;
\href{http://www.arXiv.org/abs/hep-ph/9806356}
  {\mbox{\textsf{hep-ph/9806356}}}.

\bibitem[Kanazawa \etal(2000)Kanazawa and Koike]{Kanazawa:2000hz.0}
Y.~Kanazawa and Y.~Koike, \emph{Phys. Lett.} \textbf{B478} (2000)~121;
\href{http://www.arXiv.org/abs/hep-ph/0001021}
  {\mbox{\textsf{hep-ph/0001021}}};
\addcite{Kanazawa:2000kp.0}.

\bibitem[Anselmino \etal(2003)Anselmino, D'Alesio and
  Murgia]{Anselmino:2002pd.0}
M.~Anselmino, U.~D'Alesio and F.~Murgia, \emph{Phys. Rev.} \textbf{D67} (2003)
  074010;
\href{http://www.arXiv.org/abs/hep-ph/0210371}
  {\mbox{\textsf{hep-ph/0210371}}};
\addcite{Anselmino:2002gf.0}.

\bibitem[Politzer(1974)Politzer]{Politzer:1974fr}
H.D.~Politzer, \emph{Phys. Rep.} \textbf{14} (1974)~129.

\bibitem[Ratcliffe(1999)Ratcliffe]{Ratcliffe:1998pq}
P.G.~Ratcliffe, \emph{Eur. Phys. J.} \textbf{C8} (1999)~403;
\href{http://www.arXiv.org/abs/hep-ph/9806369}
  {\mbox{\textsf{hep-ph/9806369}}}.

\bibitem[Anselmino \etal(2002)Anselmino \etal]{Anselmino:2001vn}
M.~Anselmino, V.~Barone, A.~Drago and F.~Murgia, in Proc. of the \emph{{ECT}
  Conf. on the Spin Structure of the Proton} (Trento, July 2001), eds.
  S.D.~Bass, A.~De~Roeck and A.~Deshpande;
\emph{Nucl. Phys. \volumeface{B} (Proc. Suppl.)} \textbf{105} (2002)~132;
\href{http://www.arXiv.org/abs/hep-ph/0111044}
  {\mbox{\textsf{hep-ph/0111044}}}.

\bibitem[Brodsky \etal(2002)Brodsky, Hwang and Schmidt]{Brodsky:2002cx}
S.J.~Brodsky, D.S.~Hwang and I.~Schmidt, \emph{Phys. Lett.} \textbf{B530}
  (2002)~99;
\href{http://www.arXiv.org/abs/hep-ph/0201296}
  {\mbox{\textsf{hep-ph/0201296}}}.

\bibitem[Collins(2002)Collins]{Collins:2002kn}
J.C.~Collins, \emph{Phys. Lett.} \textbf{B536} (2002)~43;
\href{http://www.arXiv.org/abs/hep-ph/0204004}
  {\mbox{\textsf{hep-ph/0204004}}}.

\additem{Ratcliffe:2002xv.0}{\ignorespaces in Proc. of the \emph{{XV} Int.
Symp. on High-Energy Spin
  Physics---{SPIN}~2002} (Long Island, Sept. 2002), eds. Y.I.~Makdisi,
  A.U.~Luccio and W.W.~Mackay;
\emph{AIP Conf. Proc.} \textbf{675} (2003)~176;
\href{http://www.arXiv.org/abs/hep-ph/0211232}
  {\mbox{\textsf{hep-ph/0211232}}}}

\additem{DeGrand:1981pe.e}{\ignorespaces \emph{erratum}, \ibid \textbf{D31}
(1985)~661}

\additem{Efremov:1982sh.t}{\ignorespaces \emph{transl.}, \emph{Sov. J. Nucl.
Phys.} \textbf{36} (1982)~140}

\additem{Efremov:1985ip.0}{\ignorespaces \emph{Phys. Lett.} \textbf{B150}
(1985)~383}

\additem{Qiu:1992wg.0}{\ignorespaces \emph{Nucl. Phys.} \textbf{B378}
(1992)~52}

\additem{Kanazawa:2000kp.0}{\ignorespaces \ibid \textbf{B490} (2000)~99;
\href{http://www.arXiv.org/abs/hep-ph/0007272}
  {\mbox{\textsf{hep-ph/0007272}}}}

\additem{Anselmino:2002gf.0}{\ignorespaces in Proc. of the \emph{{XV} Int.
Symp. on High-Energy Spin
  Physics---{SPIN}~2002} (Long Island, Sept. 2002), eds. Y.I.~Makdisi,
  A.U.~Luccio and W.W.~Mackay;
\emph{AIP Conf. Proc.} \textbf{675} (2003)~474;
\href{http://www.arXiv.org/abs/hep-ph/0211198}
  {\mbox{\textsf{hep-ph/0211198}}}}

\end{thebibliography}

\defacronym {SSA}  {single-spin asymmetry}[single-spin asymmetries]
\defacronym {DIS}  {deeply-inelastic scattering}
\defacronym*{DY}   {Drell--Yan}
\defacronym*{EM}   {electromagnetic}
\defacronym {LO}   {leading{\spaceorhyphen}order}
\defacronym {NLO}  {next-to-leading{\spaceorhyphen}order}
\defacronym*{pQCD} {perturbative QCD}
\defacronym*{QCD}  {quantum chromodynamics}
\acroforce{EM} \acroforce{QCD}
\end{document}